\documentclass[pdflatex,sn-mathphys-num]{sn-jnl}

\usepackage{graphicx}%
\usepackage{multirow}%
\usepackage{amsmath,amssymb,amsfonts}%
\usepackage{amsthm}%
\usepackage[title]{appendix}%
\usepackage{xcolor}%
\usepackage{textcomp}%
\usepackage{manyfoot}%
\usepackage{booktabs}%
\usepackage{algorithm}%
\usepackage{algorithmicx}%
\usepackage{algpseudocode}%
\usepackage{listings}%
\usepackage{multirow}   
\usepackage{subcaption} 

\theoremstyle{thmstyleone}%
\theoremstyle{thmstyletwo}%

\theoremstyle{thmstylethree}%

\begin{document}

\title[Latent Representation for MS Prediction]{Latent Representation Learning from 3D Brain MRI for Latent Analysis and Prediction in Multiple Sclerosis}


\author[1,6]{\fnm{Trinh} \sur{Ngoc Huynh}}
\author[1]{\fnm{Nguyen} \sur{Duc Kien}}
\author[2]{\fnm{Dam} \sur{Thai Ninh}}
\author[3,7]{\fnm{Nguyen} \sur{Hai Anh}}
\author[1]{\fnm{Tran} \sur{Hiep Dinh}}
\author[4]{\fnm{Manuela} \sur{Vaneckova}}
\author[5]{\fnm{Tomáš} \sur{Uher}}
\author[6,7]{\fnm{Jeroen} \sur{Van Schependom}}
\author[6]{\fnm{Stijn} \sur{Denissen}}
\author*[1]{\fnm{Tran} \sur{Quoc Long}}\email{tqlong@vnu.edu.vn}
\author[1]{\fnm{Nguyen} \sur{Linh Trung}}
\author[2,7,8]{\fnm{Guy} \sur{Nagels}}

\affil[1]{\orgdiv{VNU University of Engineering and Technology},
            \orgaddress{\city{Hanoi}, \country{Vietnam}}}
\affil[2]{\orgdiv{VNU Information Technology Institute}, 
            \orgname{Vietnam National University}, 
            \orgaddress{\city{Hanoi}, \country{Vietnam}}}
\affil[3]{\orgdiv{National Institute of Neuroscience},
           \orgname{Bach Mai Hospital}, 
           \orgaddress{\city{Hanoi}, \country{Vietnam}}}
\affil[4]{\orgdiv{Department of Radiology}, 
          \orgname{First Faculty of Medicine, Charles University, General University Hospital}, 
          \orgaddress{\city{Prague}, \country{Czech Republic}}}
\affil[5]{\orgdiv{Department of Neurology and Center of Clinical Neuroscience}, 
          \orgname{First Faculty of Medicine, Charles University, General University Hospital}, 
          \orgaddress{\city{Prague}, \country{Czech Republic}}}
\affil[6]{\orgdiv{Department of Electronics and Informatics}, 
          \orgname{Vrije Universiteit Brussel}, 
          \orgaddress{\city{Brussels}, \country{Belgium}}}
\affil[7]{\orgdiv{AIMS Lab, Center for Neurosciences}, 
          \orgname{Vrije Universiteit Brussel}, 
          \orgaddress{\city{Brussels}, \country{Belgium}}}
\affil[8]{\orgdiv{Department of Neurology}, 
          \orgname{UZ Brussel}, 
          \orgaddress{\city{Brussels}, \country{Belgium}}}


\abstract{
\textbf{Background:} Neurological diseases can lead to significant cognitive impairment. Linking MRI-derived measures of brain structure with clinical assessments of cognitive function remains challenging, as conventional statistical methods and traditional machine learning approaches may struggle to capture complex relationships between brain structure and cognitive outcomes, hindering the development of reliable imaging biomarkers. Although deep learning methods offer greater predictive capabilities, their black-box nature limits interpretability, posing a major obstacle to clinical translation.

\textbf{New method:} Latent representation learning via generative models can provide informative representations and support deeper analysis and cognitive biomarker prediction. While standard generative adversarial networks and diffusion models primarily focus on generating images from random noise, variational autoencoders offer compact, probabilistic latent representations that are well suited for downstream tasks. In this study, we propose InfoVAE-Med3D, an adaptation of the InfoVAE framework for three-dimensional brain magnetic resonance imaging(MRI) to learn informative latent representations. By explicitly relaxing the mutual information regularization penalty between inputs and latent representations, our method aims to capture richer information from input MRI data, facilitating the analysis of brain-related characteristics and downstream prediction.

\textbf{Results:} We evaluate InfoVAE-Med3D on two datasets: a healthy control dataset (\(n=6{,}503\)) containing chronological age and sex information, and a Multiple Sclerosis (MS) dataset (\(n=916\)) collected from the Multiple Sclerosis Center of the First Medical Faculty, Charles University, Prague, containing chronological age, sex, and Symbol Digit Modalities Test (SDMT) scores. The learned latent representations capture information related to sex and brain volumetric features, including grey and white matter volumes, as reflected in sex-related clustering and smooth transitions in brain volume measures in two-dimensional visualizations. Through comprehensive evaluations, we demonstrate that InfoVAE-Med3D learns more informative latent representations, especially when the mutual information regularization penalty is completely removed. In this setting, the model generally outperforms other variational autoencoder variants in both MRI reconstruction and prediction of brain age and SDMT scores using support vector regression, while remaining competitive with deep learning-based predictive models, particularly on the MS dataset.

\textbf{Conclusion:} InfoVAE-Med3D provides informative latent representations from brain MRI data of both healthy controls and patients with multiple sclerosis. By improving predictive performance and providing interpretable insights into the learned latent representations, the proposed framework offers a promising approach for advancing biomarker discovery and the analysis of cognitive impairment in neurological diseases, particularly multiple sclerosis.
}

\keywords{Latent representation learning, Variational autoencoder, Brain age, Symbol digit modalities test, Cognitive impairment, Three-dimensional brain MRI.}



\maketitle

\section{Introduction}

Cognitive impairment (CI) is a prevalent and disabling manifestation of MS, substantially affecting patients’ daily functioning and long-term prognosis~\cite{chiaravalloti2008cognitive, benedict2020cognitive}. These deficits can be subtle in onset and heterogeneous in nature, making them challenging to detect and quantify using routine clinical assessments. Commonly, CI is assessed using neuropsychological tests, such as the Symbol Digit Modalities Test (SDMT)~\cite{smith1973symbol}, which is widely used to evaluate cognitive processing speed~\cite{sumowski2018cognition}. However, repeated assessments may be affected by practice effects, while comprehensive neuropsychological evaluations can be time-consuming and resource-intensive~\cite{portaccio2010reliability}. Structural brain MRI offers a promising complementary approach for assessing brain anatomy and disease-related changes, and is widely used in neuroscience research~\cite{filippi2010contribution, mirmosayyeb2024cognitive}. Conventional statistical analyses and early machine learning approaches have been used to investigate the relationship between brain structure and cognitive function, but capturing their complex associations remains challenging~\cite{nenning2022machine}. Further analysis of three-dimensional (3D) brain MRI data is therefore expected to reveal valuable imaging biomarkers associated with neurological disorders and CI, potentially informing preventive care and early intervention.

Recent advances in artificial intelligence, particularly deep learning (DL), have driven numerous studies using brain MRI to investigate the relationship between disease-related structural changes and cognitive function, with the aim of identifying potential cognitive biomarkers~\cite{borchert2023artificial}. Among these, brain age has attracted considerable attention as an MRI-derived biomarker, as the gap between chronological age and brain-predicted age may reflect aspects of brain aging and neurodegeneration. Brain age can be estimated indirectly from MRI-derived volumetric features, including grey matter (GM) and white matter (WM) volumes, together with demographic variables such as sex, using linear regression models~\cite{denissen2022brain}, or directly from raw MRI data using DL approaches~\cite{cole2017predicting}. Sex has also been investigated as a relevant demographic feature in brain MRI-based prediction tasks. For example, a previous study used DL for sex classification and further applied transfer learning to Alzheimer’s disease classification using a large-scale brain imaging dataset comprising 85,721 samples~\cite{lu2021classification}. Other studies have explored age and sex prediction using white matter features~\cite{he2022model} or predicted age and sex from brain shape using geometric DL~\cite{besson2021geometric}. Taken together, previous studies have highlighted the value of brain age as a biomarker and the potential relevance of sex and structural brain features to cognitive biomarker prediction. However, conventional end-to-end DL models often function as black boxes that map input data directly to output labels, limiting the direct investigation of important features associated with brain structure and cognitive function~\cite{munroe2024applications}.

A promising direction is latent representation learning, which maps brain MRI scans to lower-dimensional representations that can be analyzed for brain-related characteristics and used in downstream prediction tasks. These representations provide compact encodings of the input data and can encode information relevant to brain structure and cognitive function. Notably, generative models provide a framework for learning representations of complex data by modeling their underlying distributions. Among the widely studied generative architectures are Generative Adversarial Networks (GANs)~\cite{goodfellow2014generative} and Diffusion Models (DMs)~\cite{ho2020denoising}. However, GANs may suffer from training instability and mode collapse, while diffusion models operating directly on high-dimensional MRI data can incur substantial computational costs during training and sampling. In contrast, Variational Autoencoders (VAEs)~\cite{kingma2013auto} explicitly learn probabilistic latent representations through an encoder--decoder framework, making them a natural choice for studying information encoded in lower-dimensional representations. Variants such as $\beta$-VAE~\cite{higgins2017beta} further modify the learning objective to encourage disentangled latent representations, while InfoVAE~\cite{zhao2017infovae} introduces a more flexible objective to improve the informativeness of latent representations. Motivated by these properties, we adopt the VAE family as a starting point for learning and analyzing latent representations from brain MRI and investigating their utility for cognitive biomarker prediction in MS.

\vspace{0.5em}
\noindent\textbf{Contribution: } In this paper, we propose InfoVAE-Med3D, an adaptation of the InfoVAE framework~\cite{zhao2017infovae} for learning latent representations from 3D brain MRI. We investigate the information captured by these representations through latent-space analysis and evaluate their utility for MRI reconstruction and downstream prediction. Our main contributions are fourfold: (i) adapting InfoVAE to 3D brain MRI and evaluating the effect of relaxing the mutual information regularization penalty on latent representation learning, (ii) evaluating the learned representations on downstream regression tasks, including brain age prediction and SDMT prediction using support vector regression, (iii) analyzing the learned latent representations through two-dimensional visualizations to examine their associations with sex and grey and white matter volumes, (iv) conducting systematic experiments on both HC and MS datasets, comparing VAE variants and DL models, and selecting suitable model configurations for latent representation learning and downstream prediction.

\section{Method}\label{sec2}

\begin{figure}[b!]
    \centering
    \includegraphics[width=\textwidth]{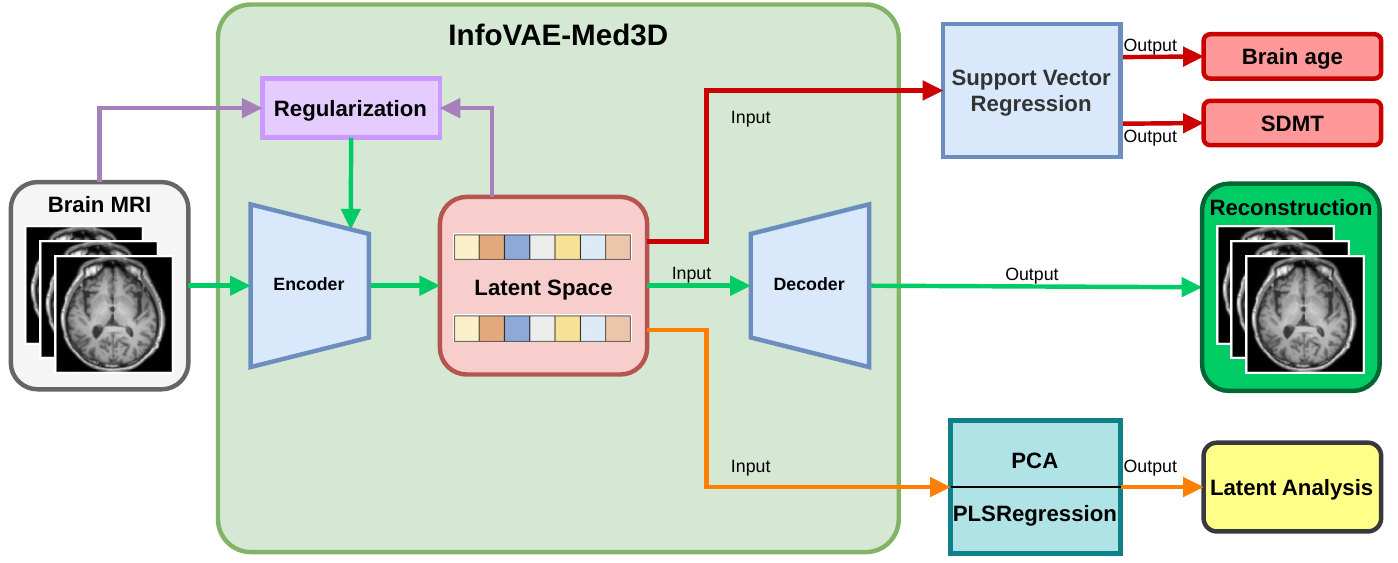}
    \caption{Overview of the proposed pipeline using InfoVAE-Med3D for latent representation learning, followed by latent-space analysis and cognitive biomarker prediction.}
    \label{fig:pipeline}
    \vspace{-0.25cm}
\end{figure}

We first revisit the standard VAE~\cite{kingma2013auto} objective and examine its extension to the InfoVAE framework~\cite{zhao2017infovae}. To study latent representation learning, we apply the InfoVAE framework to 3D brain MRI volumes and examine the role of the mutual information (MI) term in the learning objective. The learned representations are subsequently analyzed and used for downstream prediction. Formally, we consider a dataset of 3D brain MRI volumes:
\begin{equation}
    \mathcal{D} = \left\{\big(X^{(i)}, y^{(i)}\big)\right\}_{i=1}^{N},
\end{equation}
where \(N\) denotes the number of samples, \(X^{(i)} \in \mathbb{R}^{H \times W \times D}\) represents the \(i\)-th 3D brain MRI volume, with height \(H\), width \(W\), and depth \(D\), and \(y^{(i)}\) denotes the associated variables, including demographic, clinical, and brain volumetric measures, where available. These variables are not used during latent representation learning but are used, where applicable, for subsequent latent-space analysis and downstream prediction. The overall pipeline is illustrated in Figure~\ref{fig:pipeline}.

\vspace{0.5em}
\noindent\textbf{Latent Representation Learning via VAE: }  We learn latent representations by modeling the distribution of 3D brain MRI volumes without using the associated variables during training. The associated variables are reserved for subsequent latent-space analysis and downstream prediction. Let \(p_{\mathcal{D}}(X)\) denote the underlying data distribution, represented in practice by the empirical distribution of the training set. A latent-variable generative model introduces a latent variable \(Z\) with a prior distribution \(p(Z)\), typically a standard Gaussian, and defines a conditional distribution \(p_{\theta}(X \mid Z)\) parameterized by a neural network. The marginal likelihood is obtained by integrating over the latent variable \(Z\):
\begin{equation}
    p_{\theta}(X) = \int p_{\theta}(X \mid Z)\,p(Z)\,dZ.
\end{equation}
The model aims to maximize the expected log marginal likelihood over the data distribution \(\mathbb{E}_{p_{\mathcal{D}}(X)}\big[\log p_{\theta}(X)\big]\). However, directly maximizing the log marginal likelihood is generally intractable because it requires integration over the latent variable \(Z\). To address this issue, VAEs introduce an approximate posterior \(q_{\phi}(Z \mid X)\), parameterized by an encoder network, to estimate the true posterior. This variational formulation yields a tractable lower bound on the log marginal likelihood, known as the evidence lower bound (ELBO). For each MRI volume \(X\), the ELBO is defined as:
\begin{align}
    \log p_{\theta}(X) 
        &\geq \mathbb{E}_{q_{\phi}(Z \mid X)} \big[\log p_{\theta}(X \mid Z)\big] - D_{\mathrm{KL}} \big(q_{\phi}(Z \mid X)\,\|\,p(Z)\big) \\
        &= \mathcal{L}_{\mathrm{rec}}(X) - \mathcal{L}_{\mathrm{reg}}(X) \\
        &= \mathcal{L}_{\mathrm{ELBO}}(X),
\end{align}
where the inequality follows from Jensen's inequality. Here, \(\mathcal{L}_{\mathrm{rec}}(X)\) denotes the reconstruction term, which encourages accurate reconstruction of the input MRI volume, and \(\mathcal{L}_{\mathrm{reg}}(X)\) denotes the KL-divergence regularization term, which encourages the approximate posterior to remain close to the prior distribution. The ELBO provides a tractable objective for training the encoder and decoder by maximizing the lower bound on the log marginal likelihood. For the entire dataset, the ELBO objective is defined as the expectation over the empirical data distribution:
\begin{align}
    \mathcal{L}_{\mathrm{ELBO}} 
        &= \mathbb{E}_{p_{\mathcal{D}}(X)}\big[\mathcal{L}_{\mathrm{ELBO}}(X)\big] \\
        &= \mathbb{E}_{p_{\mathcal{D}}(X)}\big[\mathcal{L}_{\mathrm{rec}}(X)\big] 
        - \mathbb{E}_{p_{\mathcal{D}}(X)}\big[\mathcal{L}_{\mathrm{reg}}(X)\big].
\end{align}
This VAE formulation provides continuous, lower-dimensional latent representations suitable for downstream tasks. However, the regularization imposed by the prior may limit the informativeness of these representations.

\vspace{0.5em}
\noindent\textbf{Mutual Information Regularization: } To investigate the effect of KL regularization on the informativeness of latent representations, we adopt the InfoVAE formulation~\cite{zhao2017infovae} to examine the KL regularization term. To this end, we consider the MI between the input \(X\) and its latent representation \(Z\), which quantifies their statistical dependency under the joint distribution \(q_{\phi}(X,Z)\):
\begin{equation}
    \mathrm{MI}_{q}(X;Z) = \mathbb{E}_{q_{\phi}(X,Z)}\left[
        \log \frac{q_{\phi}(X,Z)}{q_{\phi}(X)\,q_{\phi}(Z)}
    \right].
\end{equation}
Here, \(q_{\phi}(X,Z)=p_{\mathcal{D}}(X)q_{\phi}(Z\mid X)\) denotes the joint distribution, and \(q_{\phi}(Z)\) is the aggregated posterior. Accordingly, the expected KL regularization term in the original ELBO can be expressed as:
\begin{equation}
    \mathbb{E}_{p_{\mathcal{D}}(X)}
    \big[\mathcal{L}_{\mathrm{reg}}(X)\big]
    = \mathrm{MI}_{q}(X;Z)
    + D_{\mathrm{KL}}\!\big(q_{\phi}(Z)\,\|\,p(Z)\big),
\end{equation}
This decomposition separates the mutual information between the input and latent representation from the KL divergence between the aggregated posterior and the prior, whereas the original ELBO imposes a more restrictive regularization on the approximate posterior for each input. Building on this formulation, we design InfoVAE-Med3D by reweighting these two terms to adjust the trade-off between MI and aggregated-posterior matching. The resulting objective is defined as:
\begin{align}
    \mathcal{L}_{\mathrm{InfoVAE-Med3D}}
    &= \mathbb{E}_{p_{\mathcal{D}}(X)}\!\big[\mathcal{L}_{\mathrm{rec}}(X)\big] 
    - \alpha\,\mathrm{MI}_{q}(X;Z) 
    - \beta\, D_{\mathrm{KL}}\!\big(q_{\phi}(Z)\,\|\,p(Z)\big) \label{eq:infovae_mi}\\
    &= \mathbb{E}_{p_{\mathcal{D}}(X)}\![\mathcal{L}_{\mathrm{rec}}(X)] 
    \!- \alpha\mathbb{E}_{p_{\mathcal{D}}(X)}\![\mathcal{L}_{\mathrm{reg}}(X)]
    \!- \hat\beta D_{\mathrm{KL}}\!(q_{\phi}(Z)\|p(Z)) \label{eq:infovae_implement}.
\end{align}
The two formulations are equivalent: Equation~\ref{eq:infovae_mi} explicitly highlights the role of the MI term, whereas Equation~\ref{eq:infovae_implement} is more convenient for practical implementation because it avoids direct MI estimation. The coefficient \(\alpha\) controls the penalty on the MI term as indicated by its negative sign. Larger values impose a stronger penalty, potentially suppressing latent information, whereas smaller values allow more information to be retained in the latent representations. In addition, the coefficient \(\hat{\beta}=\beta-\alpha\) controls the strength of aggregate-posterior matching to the prior. Stronger aggregate-posterior regularization encourages closer alignment between the aggregated posterior and the prior, but may affect reconstruction quality. Therefore, appropriately selecting both \(\alpha\) and \(\hat{\beta}\) is important for learning latent representations that retain informative features of the input while maintaining alignment between the aggregated posterior and the prior, potentially supporting generalization to downstream prediction tasks.

\vspace{0.5em}
\noindent\textbf{Latent-Space Analysis: } Once trained, the model's latent representations are used to investigate the information they encode and evaluate their utility in downstream tasks. Specifically, we extract a latent vector \(Z \in \mathbb{R}^{d}\) from each 3D brain MRI volume \(X \in \mathbb{R}^{H \times W \times D}\), where \(d \ll HWD\). For latent-space analysis, we apply dimensionality reduction to project the latent representations into a two-dimensional (2D) space for visualization and to examine their distribution and potential clustering patterns. We employ two statistical methods: Principal Component Analysis (PCA)~\cite{jolliffe2016principal}, an unsupervised linear method that identifies directions of maximum variance in the latent representations, and Partial Least Squares Regression (PLS Regression)~\cite{abdi2010partial}, a supervised method that identifies latent-space directions maximizing the covariance between the representations and the associated target variables. Both methods learn a projection matrix \(W \in \mathbb{R}^{d \times 2}\), mapping the latent representations into a 2D space. The 2D latent representations are then computed as:
\begin{equation}
    Z_{\mathrm{2D}} = ZW,
\end{equation}
where \(W\) is learned separately for each method. We visualize these projections and examine their associations with sex and GM/WM volumes to assess whether information related to these characteristics is encoded in the learned representations.

\vspace{0.5em}
\noindent\textbf{Cognitive Outcome Prediction: } We use the latent vector \(Z\) directly as input to Support Vector Regression (SVR)~\cite{montesinos2022support} to predict continuous target variables. Formally, a linear SVR model seeks a regression function:
\begin{equation}
    f(Z) = \langle w, Z \rangle + b,
\end{equation}
where \(w\) and \(b\) denote the weight vector and bias, respectively. SVR employs an \(\epsilon\)-insensitive loss, which defines a tolerance region of width \(2\epsilon\) around the regression function. Prediction errors within this region incur no loss, whereas errors exceeding the tolerance threshold are penalized in proportion to their magnitude beyond \(\epsilon\). This allows SVR to disregard small deviations between predicted and observed values while focusing on larger errors. To capture potentially nonlinear relationships between latent representations and target variables, SVR can also employ kernel functions, such as the radial basis function (RBF), polynomial, and sigmoid kernels. These kernels enable nonlinear regression by implicitly mapping the inputs into a feature space in which a linear regression function is learned. We use SVR to predict brain age and SDMT scores, thereby evaluating the predictive utility of the learned representations for these tasks.

\section{Experimental Setup}\label{sec3}

\subsection{Dataset}
We conduct extensive experiments on two datasets of 3D T1-weighted MRI brain scans. The first dataset, referred to as the healthy control (HC) dataset, consists of 6,503 unique healthy subjects, each contributing one MRI session, with the MRI data collected from multiple publicly available neuroimaging repositories~\cite{di2014autism,di2017enhancing,adhd2012adhd,taylor2017cambridge,aine2017multimodal,zuo2014open,lamontagne2019oasis,nooner2012nki}. The HC dataset provides gender and chronological age information across a broad age range. The second dataset, referred to as the MS dataset, is a clinical MS cohort obtained from the MS Center, First Faculty of Medicine, Charles University in Prague. It comprises 916 MS patients and 2,408 longitudinal 3D T1-weighted MRI sessions, where each patient may contribute multiple scans acquired at different time points. The available clinical and demographic variables include chronological age, SDMT scores, and gender. The MS dataset covers a relatively wide range of chronological ages and SDMT scores, providing variability for subsequent analyses. Detailed statistics of both datasets are summarized in Table~\ref{tab:datasets}.
\begin{table}[t!]
    \vspace{-0.25cm}
    \centering
    \renewcommand{\arraystretch}{1.3}
    \setlength{\tabcolsep}{15pt}
    \caption{Summary statistics of the HC and MS datasets.}
    \label{tab:datasets}
    \begin{tabular}{llcc}
        \hline
        \multicolumn{2}{l}{\textbf{Dataset}} & 
        \textbf{HC dataset} & \textbf{MS dataset} \\
        \hline
        \multicolumn{2}{l}{Subjects / Sessions} & 6,503 / 6,503 & 916 / 2,408 \\
        \multicolumn{2}{l}{Train / Val / Test} & 5,202 / 650 / 651 & 1,929 / 241 / 238 \\
        \hline
        \multirow{2}{*}{Age} 
            & Range & 18--97 & 19--74 \\
            & Mean $\pm$ SD & $43.70 \pm 21.41$ & $42.19 \pm 9.52$ \\
        \hline
        \multirow{2}{*}{SDMT} 
            & Range & -- & 16--97 \\
            & Mean $\pm$ SD & -- & $58.94 \pm 12.02$ \\
        \hline
        \multirow{2}{*}{Gender} 
            & Male & 2,971 & 731 \\
            & Female & 3,532 & 1,677 \\
        \hline
    \end{tabular}
\end{table}

The datasets were split at the subject/patient level into training, validation, and test sets, yielding an approximate 8:1:1 ratio in the number of MRI sessions. For the HC dataset, each subject contributed only one MRI session, whereas the MS dataset was split at the patient level to ensure that scans from the same patient were not distributed across different subsets. The test set was kept fixed throughout the experimental pipeline, while the training and validation sets were combined for cross-validation during downstream prediction and model selection. The resulting numbers of subjects/patients and MRI sessions in each subset are reported in Table~\ref{tab:datasets}. Overall, the datasets provide complementary characteristics for latent representation learning from 3D MRI and subsequent prediction tasks, including brain age and SDMT scores.

\subsection{Preprocessing}
To prepare the data for our experiments, we followed the preprocessing pipeline established within our research group and adopted in our previous study~\cite{ninh2025privacy}. The pipeline uses the FMRIB Software Library (FSL)~\cite{smith2004advances} and the Wood pipeline~\cite{wood2022accurate}. First, the MRI volumes were reoriented to a standard Right-Anterior-Superior coordinate system, followed by skull stripping using HD-BET and affine registration to the MNI152 template space using ANTs. The image intensities were subsequently normalized to reduce intensity variations across scans. These preprocessing steps were applied to improve spatial and intensity consistency across MRI scans and mitigate site effects, which is particularly important for the HC dataset collected from multiple imaging repositories. Finally, the preprocessed 3D T1-weighted MRI volumes were resized to a uniform spatial resolution of $128\times128\times128$ voxels and used as the input to the proposed InfoVAE-Med3D model. Additionally, since brain tissue volume measurements are not available in either dataset, SynthSeg~\cite{billot2023synthseg} is used to estimate GM and WM volumes for subsequent latent-space analysis. The reliability of these estimates was evaluated on an external dataset with ground-truth volume annotations, achieving an accuracy of up to 97\%. The estimated volumes were further filtered using a quality-control threshold of 0.65.

\subsection{Implementation}
We build the InfoVAE-Med3D architecture using 3D encoder–decoder networks from the MONAI repository~\cite{cardoso2022monai}, an open-source framework for DL in healthcare. Following the original InfoVAE formulation, maximum mean discrepancy is used to match the aggregated posterior to the prior. The model is trained using the Adam optimizer with a learning rate of \(10^{-4}\), a batch size of 2, and 300{,}000 iterations. At inference, latent representations are extracted from the encoder as 512-dimensional vectors, with this dimensionality selected based on the comparative experimental results presented in the following sections. For comparison, several VAE-based variants are implemented using the same architecture and training settings, including AE, VAE, and \(\beta\)-VAE, with \((\alpha,\hat{\beta})=(0,0), (1,0), (10,0)\), respectively. Additionally, ResNet and DenseNet baselines are implemented using the default MONAI architectures with an initial feature size of 64. Unlike the latent representation approach, these DL baselines directly predict the target variables without explicitly extracting latent representations for downstream analysis. The latent mean vectors from InfoVAE-Med3D are used as 512-dimensional features for downstream prediction. 

For cognitive biomarker prediction and latent-space analysis, the combined training and validation sets are used for model selection and fitting the latent-space analysis, respectively. SVR is selected for prediction, with hyperparameters optimized via grid search over \(C \in \{0.01, 0.1, 1, 10, 100\}\) and four kernel types (RBF, linear, polynomial, and sigmoid) using 5-fold cross-validation. The best-performing SVR model is then retrained on the combined training and validation sets and evaluated on the independent test set. Similarly, latent-space analysis is applied to the test set using PCA and PLS Regression models fitted on the combined training and validation sets.

\subsection{Evaluation}
We evaluate the learned latent representations of InfoVAE-Med3D through reconstruction and regression tasks. For reconstruction, we employ three complementary metrics: Peak Signal-to-Noise Ratio (PSNR)~\cite{huynh2008scope}, which quantifies the fidelity of the reconstructed images based on pixel-level differences; Structural Similarity Index (SSIM)~\cite{wang2004image}, which assesses structural similarity by jointly considering luminance, contrast, and local structural information; and Learned Perceptual Image Patch Similarity (LPIPS)~\cite{zhang2018unreasonable}, which evaluates perceptual similarity based on deep feature representations. For regression, Mean Absolute Error (MAE) is adopted as the primary evaluation metric, as it directly quantifies the average magnitude of prediction errors. The coefficient of determination (\(R^2\)) is additionally reported to measure the proportion of target variance explained by the predictions, while Root Mean Squared Error (RMSE) provides a complementary measure that places greater emphasis on larger prediction errors. Together, these metrics provide complementary assessments of reconstruction fidelity, perceptual similarity, and regression performance.

\section{Experimental Results}\label{sec4}

\begin{figure}[htbp]
    \centering
    \begin{subfigure}[b]{0.49\textwidth}
        \centering
        \includegraphics[width=\linewidth]{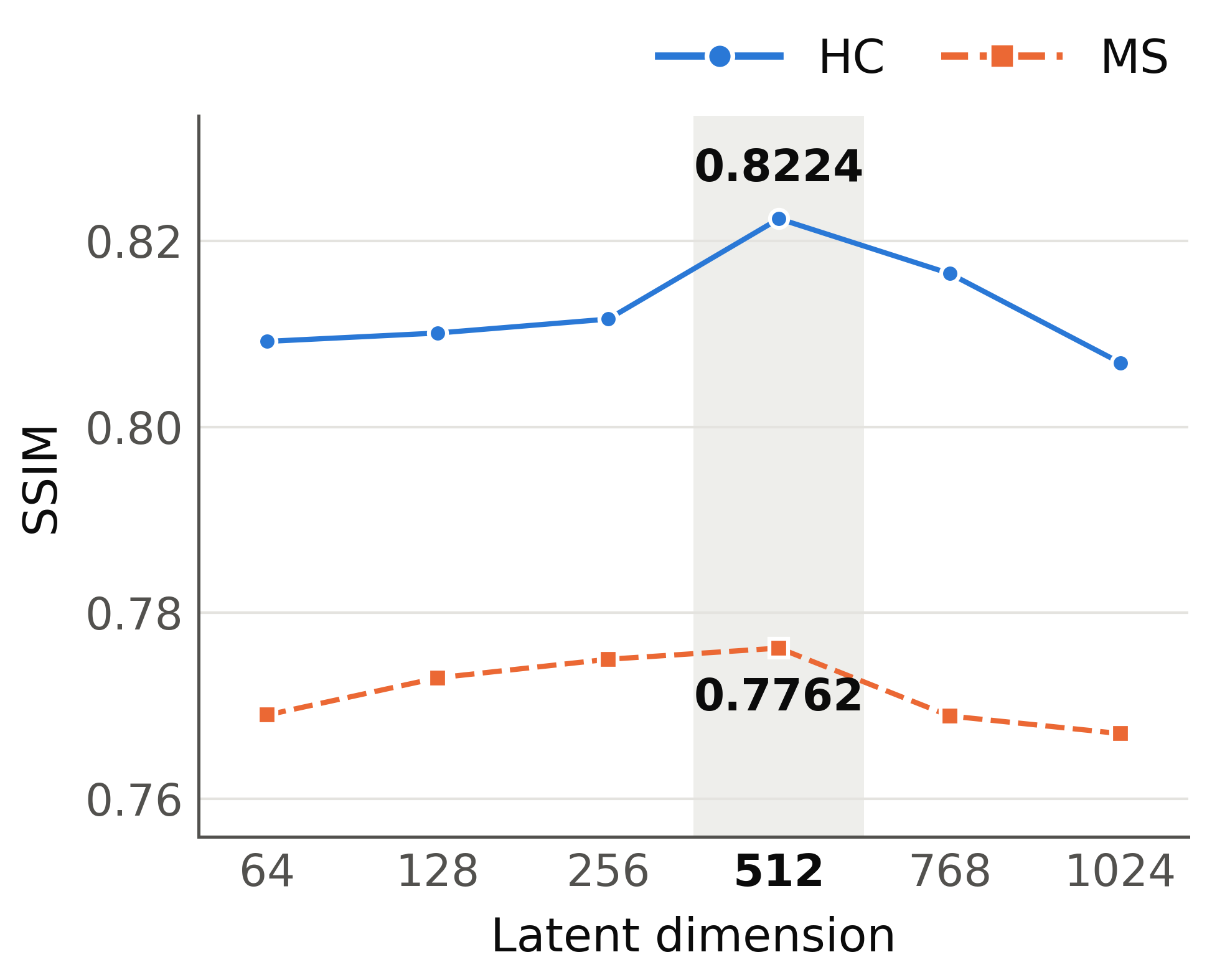}
        \subcaption{Reconstruction quality in terms of SSIM}
    \end{subfigure}
    \hfill
    \begin{subfigure}[b]{0.49\textwidth}
        \centering
        \includegraphics[width=\linewidth]{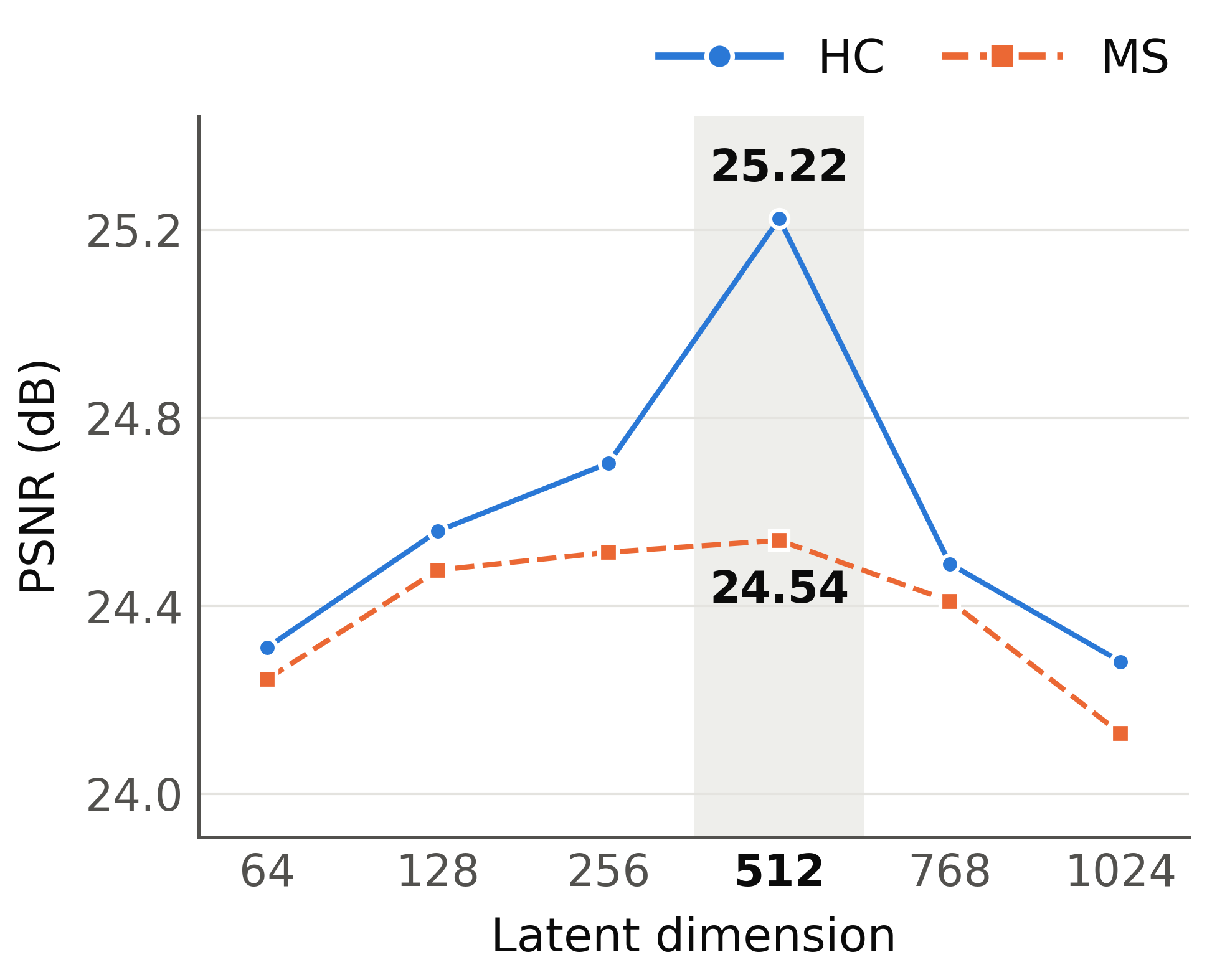}
        \subcaption{Reconstruction quality in terms of PSNR}
    \end{subfigure}
    \caption{Effect of latent dimensionality on MRI reconstruction quality.}
    \label{fig:latent_dim}
\end{figure}

\noindent\textbf{Latent Dimensionality Choices:} We investigate the effect of latent dimensionality on reconstruction quality using SSIM and PSNR, as shown in Figure~\ref{fig:latent_dim}. Both metrics exhibit a consistent trend across the HC and MS datasets, increasing as the latent dimensionality grows from 64 to 128, 256, and 512. At 512 dimensions, the model achieves its best reconstruction performance, with SSIM values of 0.8224 and 0.7762 and PSNR values of 25.22 and 24.54 dB for the HC and MS datasets, respectively. These results suggest that increasing latent dimensionality allows the model to capture more information from the input MRI volumes, thereby improving reconstruction quality. However, performance begins to decline at 768 and 1024 dimensions, suggesting that excessively high dimensionality may make the latent space more difficult to learn effectively and lead to poorer reconstruction quality. Based on these results, we select a latent dimensionality of 512 to extract latent vectors for subsequent experiments.

\noindent\textbf{Reconstruction Quality Analysis: } With a latent dimensionality of 512, Table~\ref{tab:recons} presents the reconstruction results of different InfoVAE-Med3D configurations and compares them with three baseline models: AE, VAE, and \(\beta\)-VAE. The results show that reconstruction quality is strongly influenced by the regularization parameters \(\alpha\) and \(\hat{\beta}\). With \(\hat{\beta}=1\), reducing \(\alpha\) from 1 to 0.01 improves reconstruction quality across all metrics. This suggests that weakening the MI penalty benefits reconstruction in the evaluated settings. We then set \(\alpha=0\) to investigate the effect of \(\hat{\beta}\). Increasing \(\hat{\beta}\) from 0.1 to 1 improves reconstruction quality, whereas increasing it further to 10 leads to a slight decline. The best overall configuration is \((\alpha=0,\hat{\beta}=1)\), achieving SSIM values of 0.8224 and 0.7762 and PSNR values of 25.22 and 24.54 on the HC and MS datasets, respectively. In comparison, VAE and \(\beta\)-VAE incorporate KL-based regularization, which indirectly constrains the MI between the input and latent representation. Their reconstruction performance is comparable to that of InfoVAE-Med3D with \(\alpha=1\) or \(\alpha=0.01\), where the MI penalty remains present. Compared with the AE baseline, InfoVAE-Med3D achieves higher SSIM on both datasets, while AE retains a slight advantage in LPIPS on HC and PSNR on MS. However, our focus is on learning latent representations for downstream tasks rather than solely on image compression. Overall, the configuration \((\alpha=0,\hat{\beta}=1)\) achieves the best reconstruction performance among the evaluated InfoVAE-Med3D configurations, suggesting that its latent representations encode structural information from brain MRI effectively.

\begin{table}[htbp]
    \centering
    \renewcommand{\arraystretch}{1.2}
    \setlength{\tabcolsep}{5pt}
    \caption{Reconstruction results for different configurations of the proposed model compared with baseline VAE variants. The best results are highlighted in \textbf{bold}.}
    \label{tab:recons}
    \begin{tabular}{l ccc ccc}
        \toprule
        \multirow{2}{*}{Models} 
        & \multicolumn{3}{c}{HC dataset} 
        & \multicolumn{3}{c}{MS dataset} \\
        \cmidrule(lr){2-4} \cmidrule(lr){5-7}
        & SSIM $\uparrow$ & PSNR $\uparrow$ & LPIPS $\downarrow$ 
        & SSIM $\uparrow$ & PSNR $\uparrow$ & LPIPS $\downarrow$ \\
        \midrule
        \multicolumn{7}{l}{\textbf{Baseline models}} \\
        \midrule
        AE              & 0.8065 & 24.99 & \textbf{0.1395} & 0.7370 & \textbf{24.73} & 0.1609 \\
        VAE             & 0.6269 & 20.38 & 0.2329 & 0.5825 & 20.45 & 0.2500 \\
        BetaVAE         & 0.6642    & 21.39   & 0.2156    & 0.6097 & 21.75 & 0.2362 \\
        \midrule
        \multicolumn{7}{l}{\textbf{Our proposed model: InfoVAE-Med3D}} \\
        \midrule
        $\alpha=1, \hat{\beta}=1$       
        & 0.6423 & 20.91 & 0.2251 & 0.6037 & 20.87 & 0.2484 \\
        $\alpha=0.01, \hat{\beta}=1$    
        & 0.6784 & 21.68 & 0.1986 & 0.6431 & 21.99 & 0.1819 \\
        $\alpha=0, \hat{\beta}=0.1$      
        & 0.8017 & 23.90 & 0.1562 & 0.7511 & 22.81 & 0.1427 \\
        $\alpha=0, \hat{\beta}=1$        
        & \textbf{0.8224} & \textbf{25.22} & 0.1428 
        & \textbf{0.7762} & 24.54 & \textbf{0.1305} \\
        $\alpha=0, \hat{\beta}=10$       
        & 0.8135 & 24.47 & 0.1489 & 0.7652 & 23.99 & 0.1386 \\
        \bottomrule
    \end{tabular}
    \vspace{-0.25cm}
\end{table}

\begin{figure}[htbp]
    \centering
    \includegraphics[width=\linewidth]{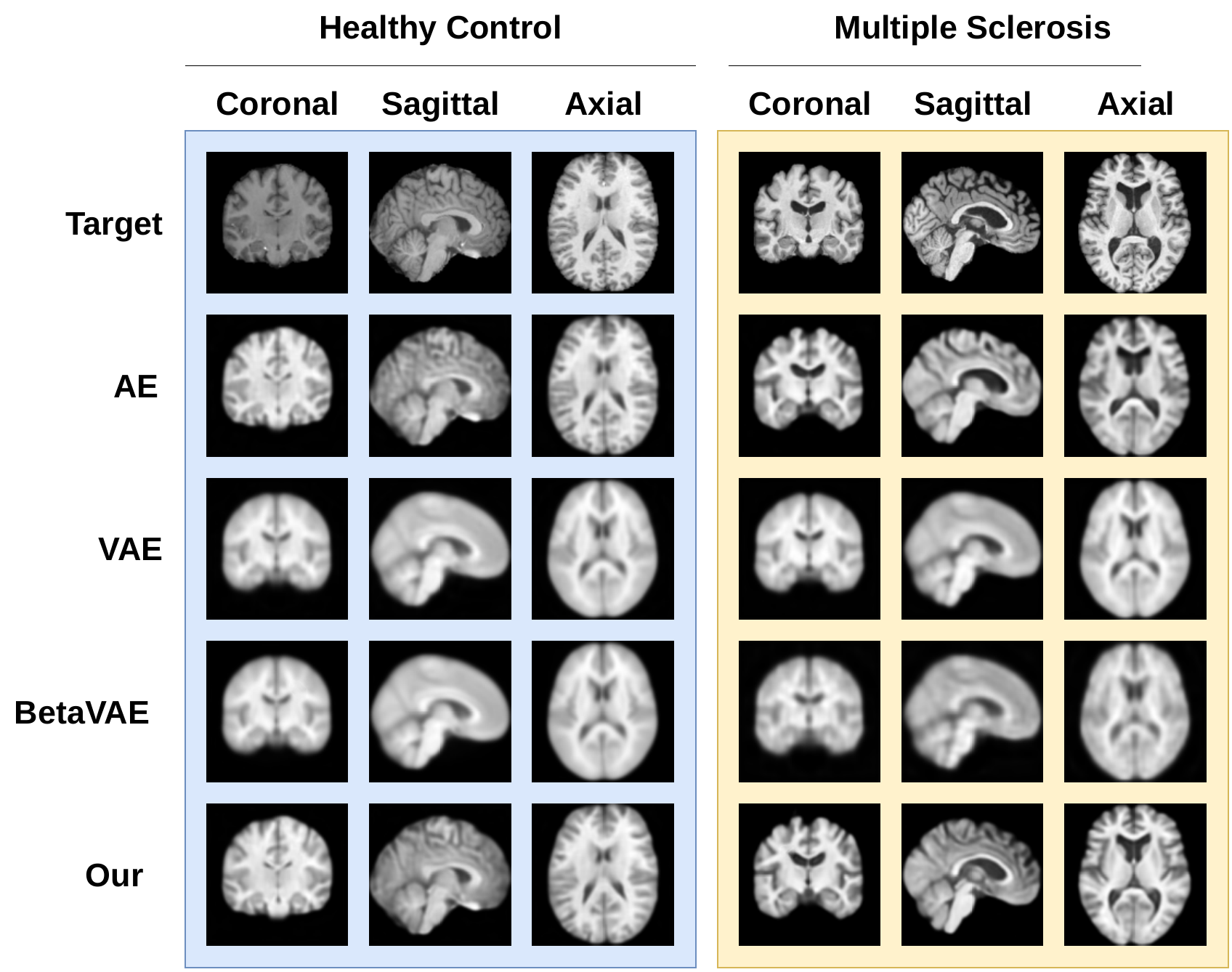}
    \caption{Qualitative reconstruction comparison of InfoVAE-Med3D and VAE baselines on two cases from the HC (in the left, blue background) and MS (in the right, yellow background) datasets.}
    \label{fig:recons}
    \vspace{-0.25cm}
\end{figure}

Furthermore, Figure~\ref{fig:recons} presents qualitative comparisons of reconstructed MRI volumes across coronal, sagittal, and axial views for representative cases from the HC and MS datasets. Blurriness is commonly observed in VAE-based models, and the results show clear differences in their ability to preserve anatomical details. VAE produces the poorest reconstructions, retaining mainly coarse outer brain structures while losing most fine details. \(\beta\)-VAE provides some improvement but still exhibits substantial blurring, particularly in regions such as the cerebellum and cortical areas. These observations are consistent with the quantitative results in Table~\ref{tab:recons}. In contrast, AE and the proposed InfoVAE-Med3D produce clearer reconstructions with richer structural details. InfoVAE-Med3D better preserves major brain structures across different anatomical views, including the separation between the cerebral hemispheres, gray and white matter patterns, and midline structures such as the brainstem. The axial and coronal views also show better preservation of inferior anatomical structures, including the cerebellum and orbital regions. Furthermore, low-intensity regions, such as the ventricles and sulcal spaces, and high-intensity white matter regions remain more distinguishable. These observations suggest that InfoVAE-Med3D better preserves anatomical structure and spatial consistency across views. Together with the quantitative results, these qualitative comparisons indicate that the proposed model learns latent representations that encode structural information from 3D brain MRI, enabling the reconstruction of the input volumes.

\begin{figure}[t!]
    \vspace{-0.25cm}
    \centering
    \begin{subfigure}[b]{\linewidth}
        \centering
        \includegraphics[width=\linewidth]{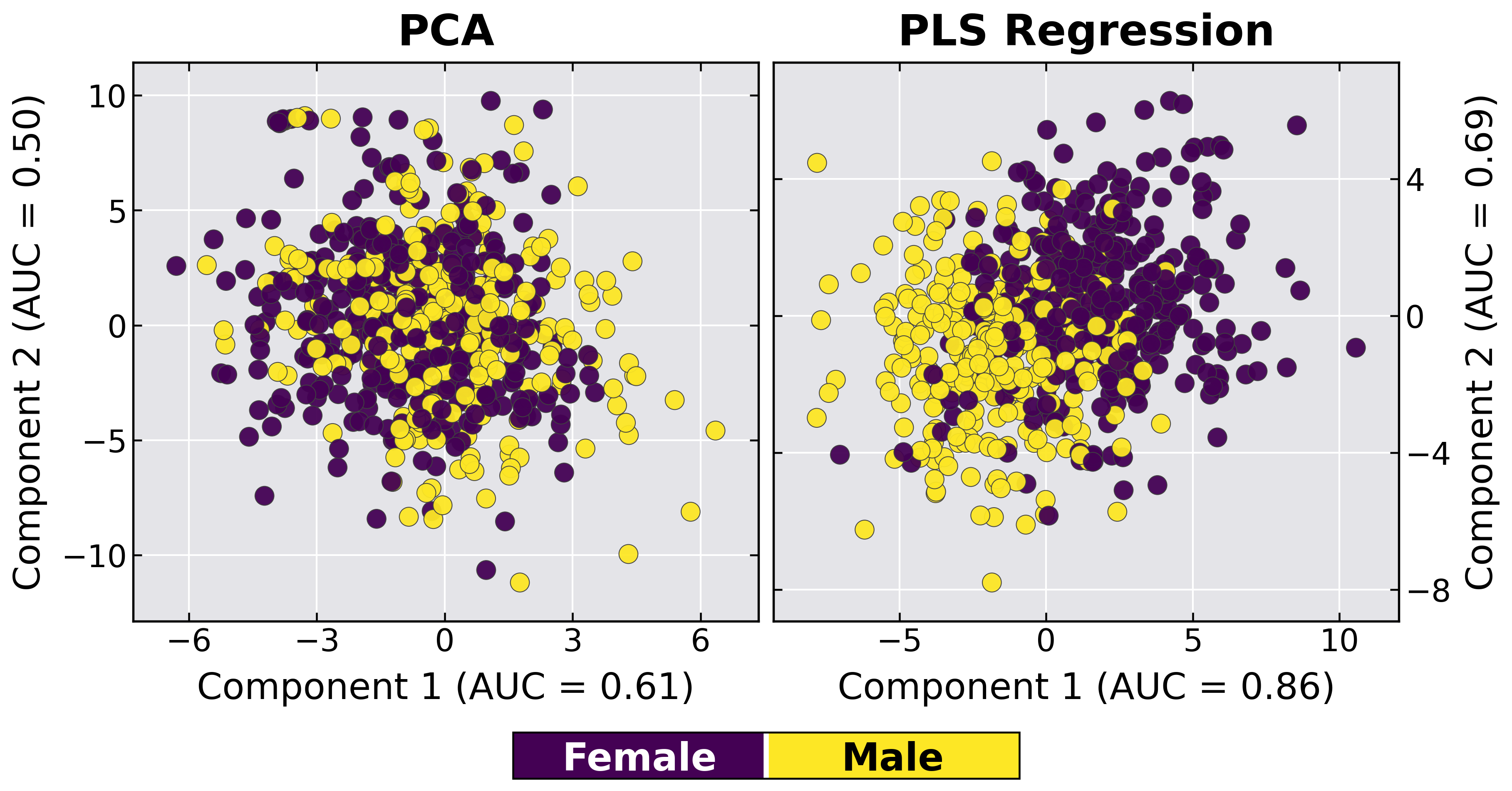}
        \subcaption{Latent representations by sex in the HC dataset.}
    \end{subfigure}
    \par\medskip
    \begin{subfigure}[b]{\linewidth}
        \centering
        \includegraphics[width=\linewidth]{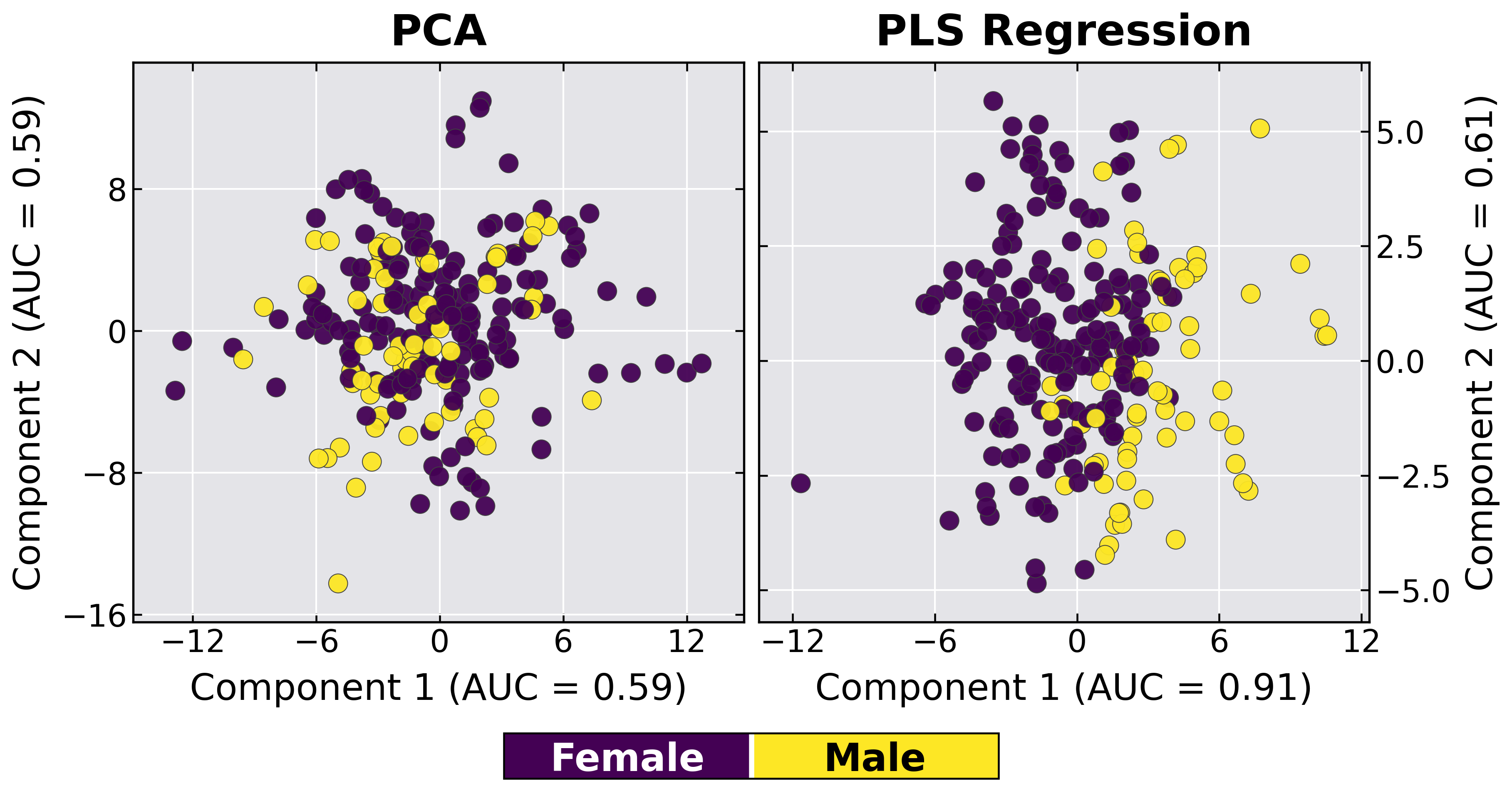}
        \subcaption{Latent representations by sex in the MS dataset.}
    \end{subfigure}
    \caption{Two-dimensional projections of latent representations colored by sex labels for the HC and MS datasets. PCA and PLS Regression projections are shown on the left and right, respectively.}
    \label{fig:latent_sex}
\end{figure}

\begin{figure}[t!]
    \vspace{-0.25cm}
    \centering
    \begin{subfigure}[b]{\linewidth}
        \centering
        \includegraphics[width=\linewidth]{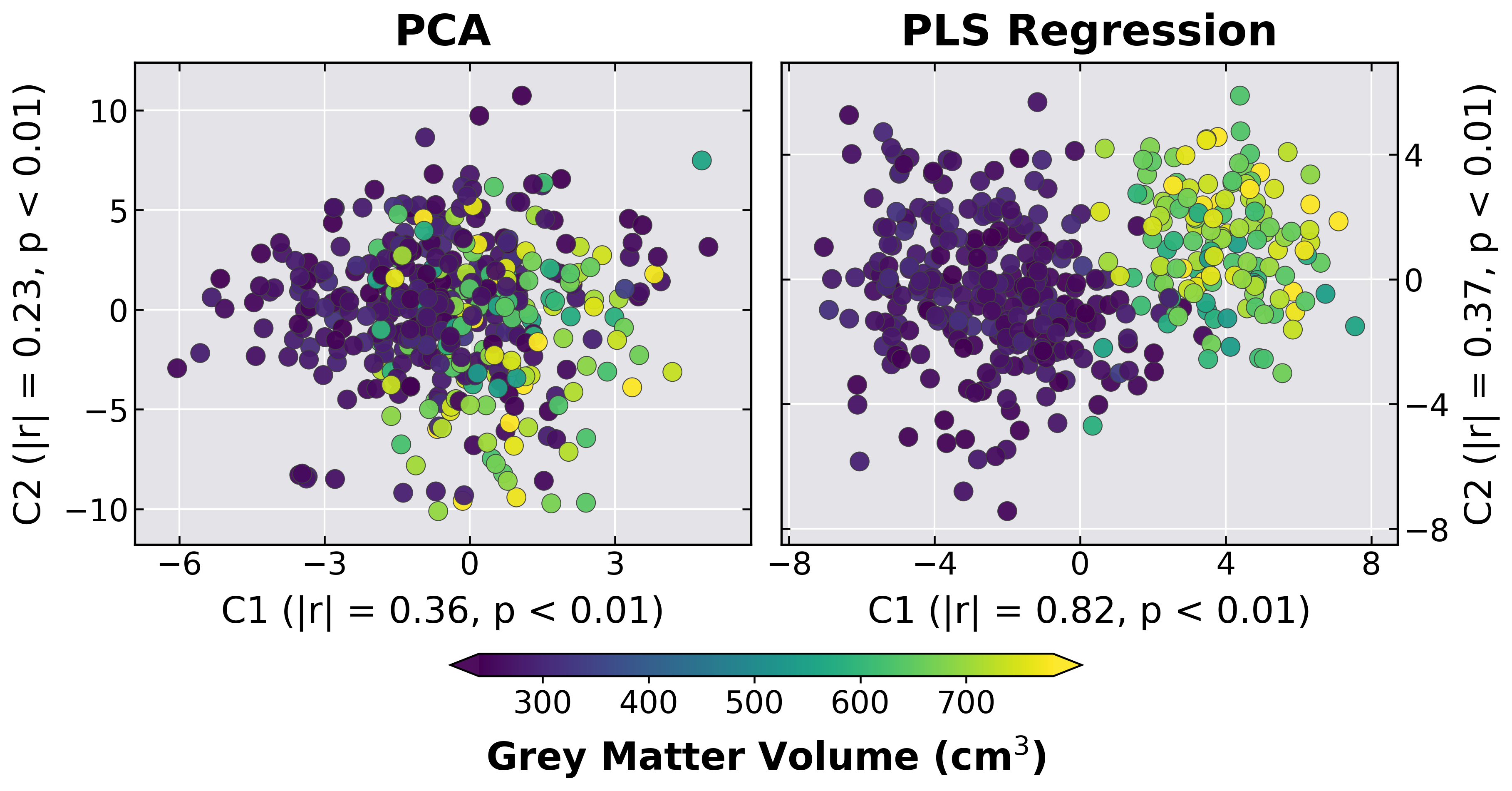}
        \subcaption{Latent representations colored by GM volume in the HC dataset.}
    \end{subfigure}
    \par\medskip
    \begin{subfigure}[b]{\linewidth}
        \centering
        \includegraphics[width=\linewidth]{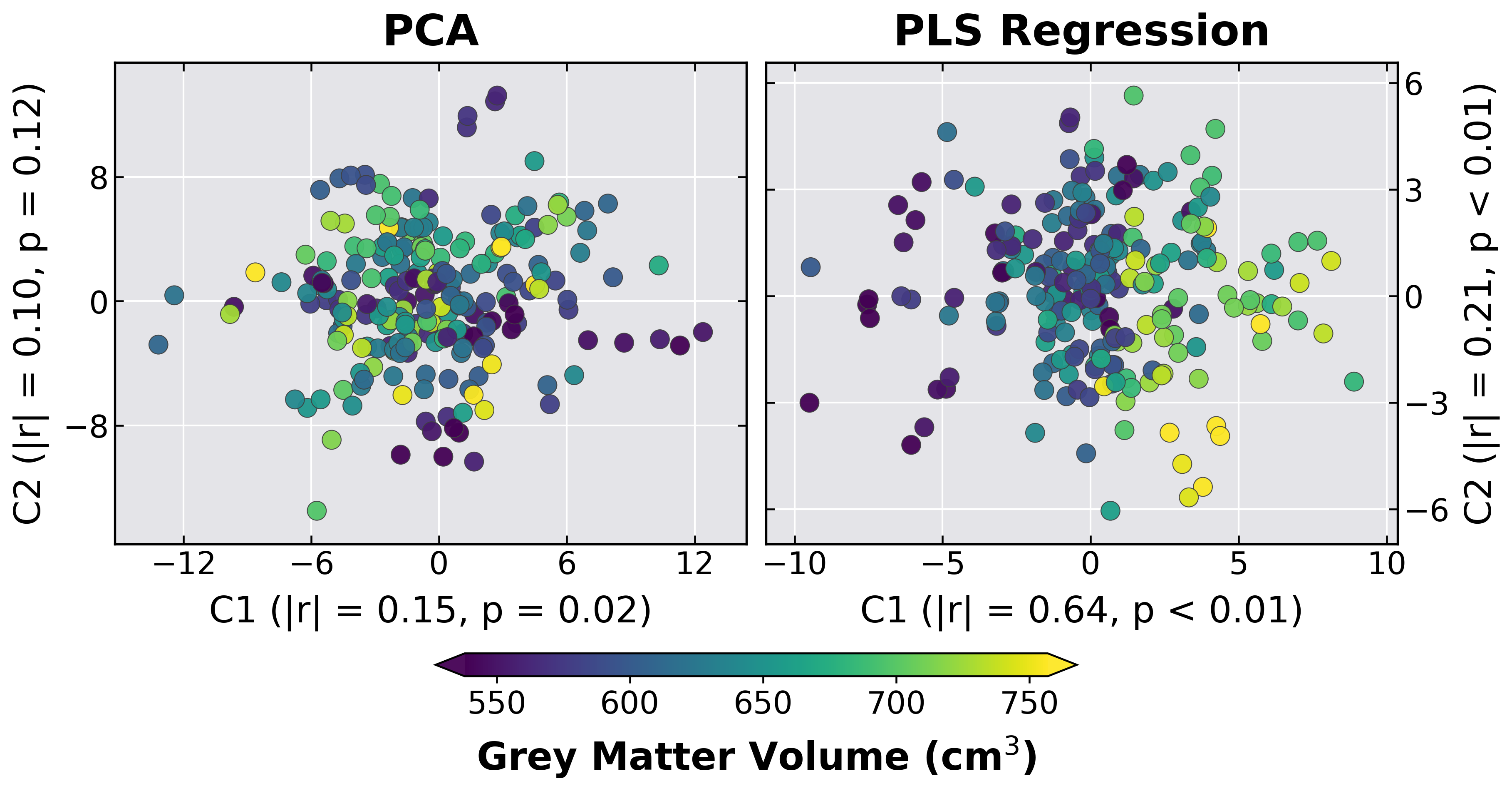}
        \subcaption{Latent representations colored by GM volume in the MS dataset.}
    \end{subfigure}
    \caption{Latent representations colored by estimated GM volumes for the HC and MS datasets. PCA and PLS Regression projections are shown on the left and right, respectively.}
    \label{fig:latent_gm}
\end{figure}

\begin{figure}[t!]
    \vspace{-0.25cm}
    \centering
    \begin{subfigure}[b]{\linewidth}
        \centering
        \includegraphics[width=\linewidth]{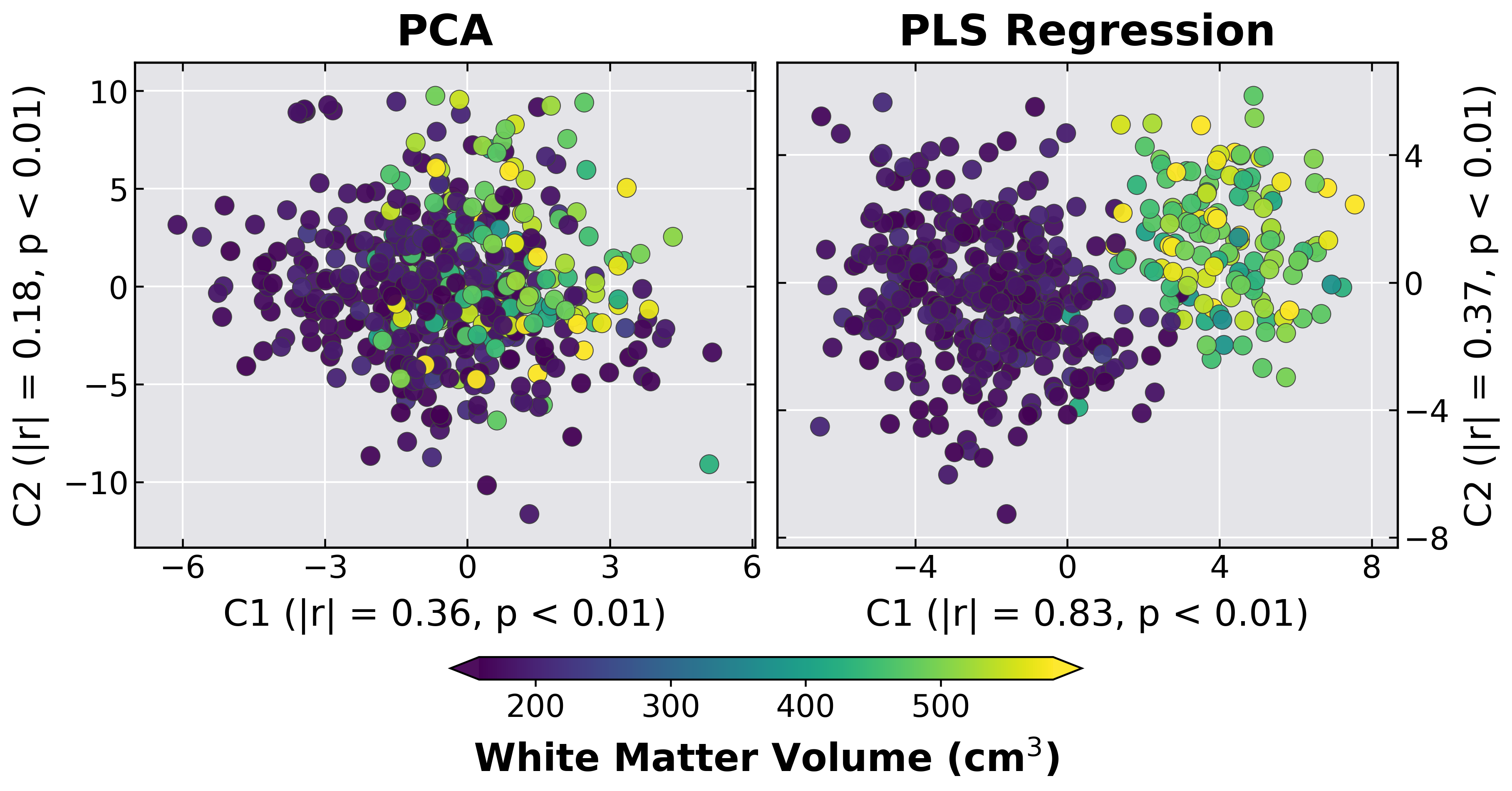}
        \subcaption{Latent representations colored by WM volume in the HC dataset.}
    \end{subfigure}
    \par\medskip
    \begin{subfigure}[b]{\linewidth}
        \centering
        \includegraphics[width=\linewidth]{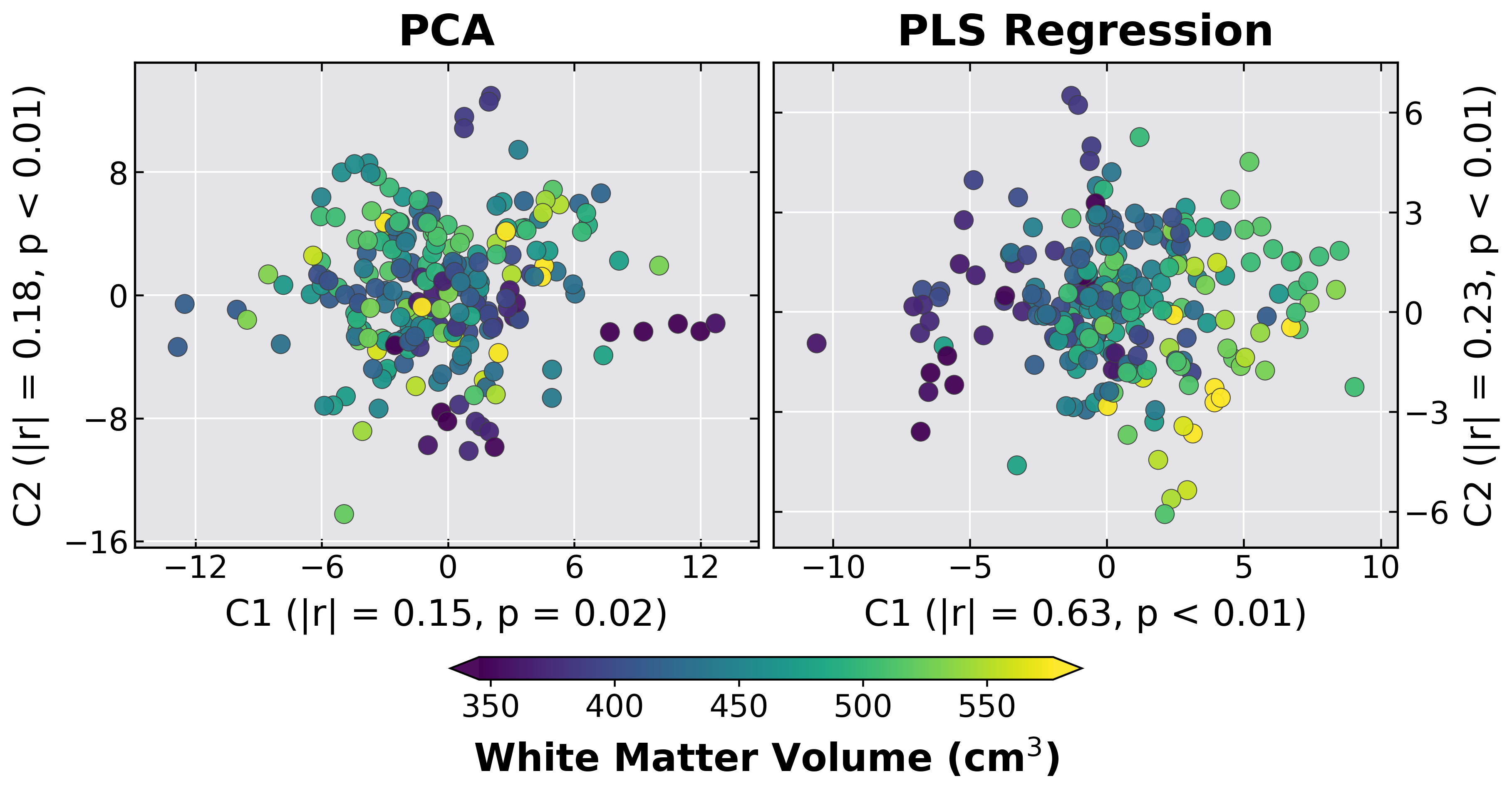}
        \subcaption{Latent representations colored by WM volume in the MS dataset.}
    \end{subfigure}
    \caption{Latent representations colored by estimated WM volumes for the HC and MS datasets. PCA and PLS Regression projections are shown on the left and right, respectively.}
    \label{fig:latent_wm}
\end{figure}

\noindent\textbf{Latent-sapce Analysis: } To gain deeper insight into the learned latent space, we examine information that has been associated with cognitive performance in previous studies. We first investigate sex-related information, with results shown in Figure~\ref{fig:latent_sex}. In both datasets, the first two principal components obtained by PCA show substantial overlap between sex groups. The corresponding AUC values are close to chance level, with the second component yielding an AUC of 0.5 on the HC dataset, indicating no discriminative ability beyond chance level. In contrast, PLS Regression yields clearer separation, with the first component achieving AUC values of 0.86 and 0.91 on the HC and MS datasets, respectively. The AUC values decrease substantially for the second component, suggesting that sex-related discriminative information is more strongly captured by the first component. As a supervised method, PLS Regression learns projections that maximize covariance with sex labels, producing latent-space representations more strongly aligned with sex-related information. These results suggest that sex-related information is encoded in the learned latent representations, although its visibility in two dimensions depends on the projection method.

Next, we investigate GM and WM volume information encoded in the latent representations from both datasets. For GM volume, as shown in Figure~\ref{fig:latent_gm}, the PCA projections reveal limited visible associations between the latent components and GM volume, particularly in the MS dataset. In contrast, the PLS Regression projection shows a clearer pattern in the HC dataset, with the first component exhibiting a strong correlation with GM volume (\(r=0.82\), \(p<0.01\)). This suggests that GM volume-related information is strongly aligned with the first PLS component. Although the transition from lower to higher GM volumes is not entirely smooth, the projected representations show a clear distinction between these ranges. In the MS dataset, the association is weaker (\(r=0.64\), \(p<0.01\)), and the pattern is less distinct. This may reflect the more complex brain structural changes associated with MS, which could make GM-related patterns more difficult to capture consistently in the learned latent representations. For WM volume, Figure~\ref{fig:latent_wm} shows a similar pattern. The PCA projections reveal limited visible associations between the latent components and WM volume, suggesting that this information is not readily captured by the first two principal components. In contrast, PLS Regression reveals a clearer separation between lower and higher WM volumes in the HC dataset, with the first component achieving a correlation of \(r=0.83\) (\(p<0.01\)). In the MS dataset, the association is weaker (\(r=0.63\), \(p<0.01\)), and the separation is less distinct. A similar trend is observed for GM volume, with weaker associations in the MS dataset, potentially reflecting the more complex brain structural changes associated with MS. Overall, these findings suggest that GM- and WM-related information is encoded in the learned latent representations, with stronger associations observed in the HC dataset than in the MS dataset. However, this information is not readily apparent in the first two PCA components but becomes more evident in the first PLS Regression component.

\begin{table}[htbp]
    \renewcommand{\arraystretch}{1.2}
    \setlength{\tabcolsep}{15pt}
    \centering
    \caption{Results of cognitive biomarker prediction using different ML models on the HC and MS datasets, reported as MAE (Mean $\pm$ SD) over 5-fold cross-validation on the combined training and validation sets. The best results are highlighted in \textbf{bold}.}
    \begin{tabular}{lccc}
        \toprule
        \multirow{2}{*}{Models} 
        & HC dataset & \multicolumn{2}{c}{MS dataset} \\
        \cmidrule(lr){2-2} \cmidrule(lr){3-4}
        & Brain age & Brain age & SDMT\\
        \midrule
        RF         & 8.192 $\pm$ 0.304 & 6.145 $\pm$ 0.368 & 8.821 $\pm$ 0.384 \\
        LR         & 6.715 $\pm$ 0.351  & 6.314 $\pm$ 0.386 & 10.648 $\pm$ 0.486 \\
        KNN        & 8.829 $\pm$ 0.336 & 7.313 $\pm$ 0.535 & 9.713 $\pm$ 0.491 \\
        XGBoost        & 7.234 $\pm$ 0.345 & 5.929 $\pm$ 0.420 & 8.853 $\pm$ 0.441 \\
        Adaboost   & 7.707 $\pm$ 0.261 & 6.055 $\pm$ 0.439 & 8.877 $\pm$ 0.459 \\
        Bagging    & 8.195 $\pm$ 0.292 & 6.123 $\pm$ 0.372 & 8.787 $\pm$ 0.428 \\
        SVR        & \textbf{6.553 $\pm$ 0.199} & \textbf{5.788 $\pm$ 0.341} & \textbf{8.670 $\pm$ 0.453} \\
        \bottomrule
    \end{tabular}
    \label{tab:train_result}
    \vspace{-0.25cm}
\end{table}

\noindent\textbf{Cognitive Outcome Prediction: } Having explored the information encoded in the latent space, we evaluate the predictive utility of the learned latent representations for brain age and SDMT score prediction using several ML models. Five-fold cross-validation is performed on the combined training and validation sets to identify a suitable model for each prediction task. As shown in Table~\ref{tab:train_result}, the relative performance of the models varies across the three prediction settings, although SVR consistently achieves the lowest mean MAE. For brain age prediction on the HC dataset, SVR achieves an MAE of \(6.553 \pm 0.199\), slightly outperforming LR, which yields an MAE of \(6.715 \pm 0.351\). The difference in mean MAE is \(0.162\), indicating a modest improvement over the next-best model. A similar pattern is observed for brain age prediction on the MS dataset, where SVR achieves an MAE of \(5.788 \pm 0.341\), compared with \(5.929 \pm 0.420\) for XGBoost. Although SVR again yields the lowest mean MAE, the improvement is relatively small (\(0.141\)). For SDMT prediction, SVR also achieves the lowest mean MAE (\(8.670 \pm 0.453\)), followed by Bagging at \(8.787 \pm 0.428\), corresponding to a difference of \(0.117\). Across all three settings, the standard deviations indicate variation in performance across the five cross-validation folds. These results show that SVR provides the best average predictive performance among the evaluated models, but the relatively small differences in mean MAE do not, by themselves, establish a statistically significant advantage. Nevertheless, its consistently lowest mean MAE across the three tasks supports selecting SVR for subsequent downstream prediction experiments using the learned latent representations.

\begin{table}[htbp]
    \centering
     \caption{Quantitative results of cognitive biomarker prediction on the HC and MS datasets, comparing latent representations learned by our proposed model, representation-learning baselines, and DL models. The best results are highlighted in \textbf{bold}.}
    \label{tab:test_result}
    \renewcommand{\arraystretch}{1.2}
    \setlength{\tabcolsep}{3pt}
    \begin{tabular}{lccccccccc}
        \toprule
        \multirow{3}{*}{Models} 
         & \multicolumn{3}{c}{HC dataset} & \multicolumn{6}{c}{MS dataset} \\
        \cmidrule(lr){2-4} \cmidrule(lr){5-10}
         & \multicolumn{3}{c}{Brain age} & \multicolumn{3}{c}{Brain age} & \multicolumn{3}{c}{SDMT} \\
        \cmidrule(lr){2-4} \cmidrule(lr){5-7} \cmidrule(lr){8-10}
         & MAE $\downarrow$ & \(R^2 \uparrow\) & RMSE $\downarrow$ 
         & MAE $\downarrow$ & \(R^2 \uparrow\) & RMSE $\downarrow$ 
         & MAE $\downarrow$ & \(R^2 \uparrow\) & RMSE $\downarrow$ \\
        \midrule
        Average             & 10.39 & 0.000 & 11.98 & 7.153 & 0.000 & 9.129 & 9.650 & -2.274 & 11.96 \\

        \midrule
        \multicolumn{10}{l}{\textbf{DL models}} \\
        \midrule
        ResNet              & 6.484 & 0.541 & 8.221 & 5.691 & 0.406 & 7.112 & 9.132 & 0.095 & 11.42 \\ 
        DenseNet            & \textbf{6.063} & \textbf{0.574} & \textbf{7.901} & 5.544 & 0.413 & 6.978 & 9.023 & 0.108 & 11.35 \\
        
        \midrule
        \multicolumn{10}{l}{\textbf{Representation-learning baselines}} \\
        \midrule
        AE                  & 7.346 & 0.422 & 9.112 & 5.704 & 0.399 & 7.078 & 9.220 & 0.056 & 11.62 \\
        VAE                 & 7.862 & 0.201 & 10.71 & 7.094 & 0.010 & 9.081 & 9.622 & 0.020 & 11.84 \\
        \(\beta\)-VAE       & 7.520 & 0.305 & 9.782 & 6.309 & 0.251 & 7.901 & 9.252 & 0.051 & 11.65 \\

        \midrule
        \multicolumn{10}{l}{\textbf{Our proposed model, InfoVAE-Med3D}} \\
        \midrule
        $\alpha=1, \hat{\beta}=1$       & 7.509 & 0.410 & 9.204 & 6.103 & 0.296 & 7.452 & 9.136 & 0.060 & 11.60 \\
        $\alpha=0, \hat{\beta}=0.1$     & 7.254 & 0.470 & 8.719 & 5.688 & 0.409 & 7.010 & 8.975 & 0.088 & 11.42 \\
        $\alpha=0, \hat{\beta}=1$       & 7.069 & 0.480 & 8.639 & \textbf{5.520} & \textbf{0.422} & \textbf{6.936} & \textbf{8.923} & \textbf{0.111} & \textbf{11.28} \\
        \bottomrule
    \end{tabular}
    \vspace{-0.25cm}
\end{table}

With SVR selected as the prediction model, we evaluate the downstream predictive performance of InfoVAE-Med3D on the test set and compare it with representation-learning baselines using their learned latent representations. We also include DL models that directly predict from 3D MRI scans. Table~\ref{tab:test_result} summarizes the results across the three prediction settings, with the training-set mean serving as a naive baseline in the first row. Among the representation-learning methods, standard VAE and \(\beta\)-VAE generally perform worse than AE and InfoVAE-Med3D. In particular, standard VAE achieves \(R^2\) values of only 0.010 for brain age prediction and 0.020 for SDMT prediction on the MS dataset, indicating limited predictive performance in these settings. The two additional configurations of InfoVAE-Med3D further assess the effect of different model settings on downstream prediction. Overall, the configuration with \(\alpha=0\) and \(\hat{\beta}=1\) achieves the best predictive performance among the evaluated InfoVAE-Med3D configurations across all three tasks. Compared with AE, this configuration reduces MAE from 7.346 to 7.069 for brain age prediction on the HC dataset, from 5.704 to 5.520 for brain age prediction on the MS dataset, and from 9.220 to 8.923 for SDMT prediction. The corresponding \(R^2\) values also increase from 0.422 to 0.480, from 0.399 to 0.422, and from 0.056 to 0.111, respectively. These results indicate that the learned latent representations provide improved predictive performance over the AE baseline across all three settings. Compared with direct DL-based prediction, InfoVAE-Med3D achieves the best results on the MS dataset for both brain age and SDMT prediction, although the differences from DenseNet are modest. On the HC dataset, however, DenseNet performs better, achieving an MAE of 6.063 compared with 7.069 for the best-performing InfoVAE-Med3D configuration. The RMSE results follow a similar overall pattern to MAE. Interestingly, brain age prediction is more accurate on the MS dataset than on the HC dataset in terms of MAE and \(R^2\). This difference may partly reflect the greater heterogeneity of the multi-source HC dataset, whereas the MS data were collected at a single site. Taken together, these findings show that InfoVAE-Med3D improves upon the representation-learning baselines and achieves competitive downstream predictive performance compared with DL models, particularly on the MS dataset. Nevertheless, the relatively low \(R^2\) of 0.111 for SDMT prediction across all models suggests that SDMT scores are difficult to predict from MRI data, potentially because test performance is sensitive to various factors that may affect individuals during the assessment.

\section{Conclusion}\label{sec5}

In this study, we proposed InfoVAE-Med3D, a latent representation learning approach for latent-space analysis and cognitive biomarker prediction in MS. By relaxing the MI regularization penalty between the input MRI volumes and their latent representations, the proposed model learn more information in the learned representations for subsequent analysis and prediction. Through PLS regression, the first component for sex discrimination achieved a high AUC, exceeding 0.9 on the MS dataset, while the first components for GM and WM volumes showed strong, statistically significant correlations with their respective targets, exceeding 0.8 on the HC dataset. Both patterns were clearly reflected in their respective two-dimensional projections. Furthermore, the learned representations were used for brain age and SDMT prediction with SVR, while PLS regression revealed associations that provide insights into the information relevant to these predictions. InfoVAE-Med3D outperforms the evaluated VAE variants in reconstruction and prediction performance, while achieving competitive results compared with DL models that directly predict from MRI volumes, particularly on the MS dataset. These findings highlight the potential of InfoVAE-Med3D for analyzing information encoded in latent representations and predicting cognitive outcomes from 3D brain MRI.
 
\vspace{0.5em}
\noindent\textbf{Limitation \& Future Work:} This study represents an initial step toward latent-space analysis and cognitive biomarker prediction in MS. Despite promising results, limitations remain, including reconstruction quality, the scope of clinical applications, and the exclusive use of T1-weighted MRI. Future work will explore advanced GenAI models to improve latent representations and enable controlled latent-space manipulation for enhanced interpretability. Additionally, incorporating FLAIR and multimodal MRI data may yield more comprehensive representations, improve biomarker prediction, and provide deeper insights into disease-related structural changes.

\section*{Acknowledgements}
Trinh Ngoc Huynh, Nguyen Duc Kien, Tran Quoc Long, and Nguyen Linh Trung were supported by the joint VUB--UET research project entitled ``Leveraging Generative Models-Derived MRI Latent Space Features for Cognitive Deterioration Prediction''. Trinh Ngoc Huynh's PhD scholarship was supported by the Research Foundation Flanders (FWO) under Senior Research Project No. G087824N. Guy Nagels is a Senior Clinical Research Fellow of the FWO under grant number 1805620N.

\bibliography{sn-bibliography}

\end{document}